\documentclass[pre,aps,twocolumn,superscriptaddress,amsmath,showpacs,floatfix]{revtex4}
\usepackage{graphicx}
\begin{document}
\title{Mechanical surface tension governs membrane thermal fluctuations}
\author{Oded Farago}
\affiliation{Department of Biomedical Engineering and Ilse
Katz Institute for Nanoscale Science and Technology, Ben Gurion University,
Be'er Sheva 84105, Israel}
\begin{abstract}
Motivated by the still ongoing debate about the various possible
meanings of the term surface tension of bilayer membranes, we present
here a detailed discussion that explains the differences between the
``intrinsic'', ``renormalized'', and ``mechanical'' tensions. We use
analytical considerations and computer simulations to show that the
membrane spectrum of thermal fluctuations is governed by the
mechanical and not the intrinsic tension. Our study highlights the
fact that the commonly used quadratic approximation of Helfrich
effective Hamiltonian is not rotationally invariant. We demonstrate
that this non-physical feature leads to a calculated mechanical
tension that differs dramatically from the correct mechanical
tension. Specifically, our results suggest that the mechanical and
intrinsic tensions vanish simultaneously, which contradicts recent 
theoretical predictions derived for the approximated Hamiltonian.
\end{abstract}
\maketitle

\section{Introduction}
\label{sec:intro}

Bilayer membranes are quasi-two-dimensional (2D) fluid sheets formed
by spontaneous self-assembly of lipid molecules in water
\cite{israelachvili}. Their elasticity is traditionally studied in the
framework of the Helfrich effective surface Hamiltonian for 2D
manifolds with local principle curvatures $c_1$ and $c_2$
\cite{helfrich:73}
\begin{equation}
{\cal H}_0=\int_{A} dS
\left[\sigma_0+\frac{1}{2}\kappa_0
\left(c_1+c_2-2c_{0}\right)^2+{\bar{\kappa}}_0c_1c_2\right],
\label{eq:helfhamiltonian}
\end{equation}
where the integration is carried over the whole surface of the
membrane. The Helfrich Hamiltonian involves four parameters: the
spontaneous curvature $c_0$, the surface tension $\sigma_0$, the
bending modulus $\kappa_0$, and the saddle-splay modulus
$\bar{\kappa}_0$. For symmetric bilayer membranes, $c_0=0$. If, in
addition, the discussion is limited to deformations that preserve the
topology of the membrane, then (by virtue of the Gauss-Bonnet theorem)
the total energy associated with the last term is a constant, and one
arrives to the more simple form
\begin{equation}
{\cal H}_0=\int_{A} dS
\left[\sigma_0+\frac{1}{2}\kappa_0\left(c_1+c_2\right)^2\right]
=\sigma_0 A+\frac{1}{2}\kappa_0 J^2,
\label{eq:helfhamiltonian2}
\end{equation}
where $A$ is the total area of the membrane and $J$, defined by
$J^2=\int dS \left(c_1+c_2\right)^2$, is the integrated total
curvature.

The surface tension appearing in Eq.(\ref{eq:helfhamiltonian2}) is
known as the ``intrinsic tension''. It represents the elastic energy
required to increase the surface of the membrane by a unit area, and
can be identified with the derivative of the energy with respect to
the area $A$, $\sigma_0=(\partial {\cal H}/\partial A)$, at constant
$J$. As this quantity is not directly measurable, its physical meaning
is still a matter of a fierce debate.  In molecular simulations, one
can attach the membrane to a ``frame'' and measure the ``mechanical
(frame) tension'', $\tau$, which is the lateral force per unit length
exerted on the boundaries of the membrane
\cite{goets:98,lindahl:00,grace:05,neder:10}. Experimentally, the
mechanical tension is routinely measured by micropipette aspiration of
vesicles \cite{rawicz:90,rawicz:00}. Formally, the mechanical tension
is obtained by taking the full derivative of the {\em free energy}\/
$F$ with respect to the frame (projected) area: $\tau=dF/dA_p$
\cite{remark1}. From a comparison of the above definitions of
$\sigma_0$ and $\tau$, it becomes clear that the intrinsic and
mechanical tensions are different quantities. Contrary the former, the
latter is a thermodynamic quantity that also depends on the entropy of
the membrane. Bilayer membranes usually exhibit relatively large
thermal undulations at room temperature \cite{safran} and, indeed,
their mechanical tension also includes an entropic contribution due to
the suppression of the amplitude of the undulations upon increasing
the projected area.

The surface tension can be also measured indirectly by recording and
analyzing the statistics of the membrane height fluctuations
\cite{duwe:90,dobereiner:03}. The analysis is based on the so called
Monge parametrization, where the surface of the fluctuating membrane
is represented by a height function, $h(x,y)$, above the frame $(x,y)$
plane. The Helfrich Hamiltonian does not have a simple form when
expressed in terms of $h(x,y)$. However, for a nearly flat membrane,
i.e., when the derivatives of $h$ with respect to $x$ and $y$ are
small ($|\partial_x h|,\ |\partial_y h|\ll 1$), one obtains the
quadratic approximation
\begin{equation}
{\cal H}^M_2=\sigma_0 A_p+\int  dxdy\left[\frac{\sigma_0}{2}
\left(\nabla h\right)^2+\frac{\kappa_0}{2}\left(\nabla^2h\right)^2\right].
\label{eq:mongehamiltonian}
\end{equation}
Note that unlike Eq.(\ref{eq:helfhamiltonian2}), the integral in
Eq.(\ref{eq:mongehamiltonian}) runs over the frame area rather than
over the area of the manifold. The quadratic approximation can be
diagonalized by introducing the Fourier transformation:
$h_q=(1/A_p)\int dxdy\, h(x,y)\exp(-i\vec{q}\cdot\vec{r})$. In Fourier
space the Hamiltonian reads
\begin{equation}
{\cal H}^M_2=\sigma_0 A_p+\frac{A_p}{2}\sum_{\vec{q}}\left(\sigma_0
q^2+\kappa_0 q^4\right) |h_q|^2,
\label{eq:fourierhamiltonian}
\end{equation}
and by invoking the equipartition theorem, we find that the mean
square amplitude of mode $\vec{q}$ (``spectral intensity''):
\begin{equation}
\langle |h_q|^2\rangle=\frac{k_BT}{A_p(\sigma_0 q^2+\kappa_0 q^4)}.
\label{eq:equipart}
\end{equation}
From the last result, it seems as if $\sigma_0$ can be extracted from
the fluctuation spectrum of the membrane.  Results of both
fully-atomistic and coarse-grained simulation
\cite{marrink:01,farago:03,cooke:05,wang:05,boek:05} show that spectral
intensity can indeed be fitted to the form
\begin{equation}
\langle |h_q|^2\rangle =\frac{k_BT}{A_p(r q^2+{\cal O}(q^4))}
\label{eq:q2coef}
\end{equation}
However, the derivation of Eq.(\ref{eq:equipart}) is based on the
approximated Hamiltonian ${\cal H}^M_2$ and, therefore, it is not
a-priory clear why the so called ``$q^2$-coefficient'' appearing in
Eq.(\ref{eq:q2coef}), $r=\sigma_0$. In fact, some theoretical studies
have argued that the $q^2$ coefficient is actually equal to the
mechanical tension $\tau$
\cite{cai:94,farago_pincus:04,schmid:11}. This conclusion has been
rejected more recently in favor of the more common interpretation that
$r=\sigma_0$ \cite{imparato:06,fournier:08}. Membrane simulations
carried at fixed mechanical tension (usually performed for $\tau=0$)
tend to agree with the result that $r=\tau$
\cite{lindahl:00,marrink:01,cooke:05,wang:05,farago:08}, but
simulations that show the opposite $r\neq\tau$ also exist
\cite{imparato:06,stecki:06}. In this paper we settle this problem and
prove, analytically and computationally, that the correct result is
indeed $r=\tau$.

\section{What does the intrinsic tension represent?}

Much of the confusion associated with the physical meaning of surface
tension in membranes is related to fact that the concept of surface
tension has been originally defined for an interface between bulk
phases (e.g., between water and oil) \cite{rowlinson_widom}. In its
original context, the surface tension represents the access free
energy $\Delta F$ per unit area $A$ of the interface between the bulk
phases
\begin{equation}
\gamma=\frac{\Delta F}{A}.
\label{eq:bulktension}
\end{equation}
When the concept of surface tension is introduced into the theory of
bilayer membranes, its meaning is distorted due to the following two
major differences between membranes and interfaces of bulk phases
\begin{enumerate}
\item In the case of bulk phases, it is often assumed that the
interface between them is flat. There is usually very little interest
in the thermal roughness of the interface, unless it is very soft. In
contrast, bilayer membranes are treated as highly fluctuating surfaces
whose elastic response is very much influenced by the entropy
associated with the thermal fluctuations.
\item The changes in the areas of both systems arise from very
different origins. In the case of, say, a water-oil interface, the
changes in the interfacial area are {\em not} produced by elastic
deformations (dilation/compression) that modify the molecular
densities of the bulk phases. Instead, they result from transfer of
molecules between the bulk phases and the interface occurring, for
instance, when the shape of the container is changed. The surface
tension $\gamma$ is essentially a chemical potential which is directly
related to the exchange parameter between the coexisting phases
\cite{dill}. The case of bilayer membranes is completely
different. Here, there is an exchange of water between the bulk fluid
and interfacial region, but almost no exchange of {\em lipid}\/
material because the concentration of free lipids in the embedding
solution is extremely low ($10^{-6}-10^{-10}$ M
\cite{israelachvili}). In other words, there is no reservoir of lipids
outside of the bilayer and, therefore, a changes in the bilayer area
result in a change in the area density of the lipids. The membrane
surface tension measures the response to this elastic deformation,
including the indirect contribution due to the exchange of water
between the bilayer and solution resulting from the deformation (which
has influence on the effective elastic moduli).
\end{enumerate}

Most earlier theoretical investigations of membranes involved the
assumption that the area per lipid $a$ is constant. This assumption
relies on the observation that the energy cost involved in density
fluctuations is much larger than the energy scale associated with
curvature fluctuations. Further assuming that the lipids are insoluble
in water (and, therefore, they all reside on the membrane where their
number $N$ is constant) implies that the total area of the membrane
$A=Na$ is constant. If that is the case, then why does one need to
include this constant in the Helfrich Hamiltonian [first term on the
right hand side of Eq.(\ref{eq:helfhamiltonian2})], and what does the
coefficient, $\sigma_0$, represent? The answer is simple. It is
technically very hard to calculate analytically the partition function
for a fluctuating manifold with a fixed area $A$. The first term in
Eq.(\ref{eq:helfhamiltonian2}) is a Lagrange multiplier that fixes the
mean area $\langle A\rangle$ of the membrane, and the value of
$\sigma_0$ is set by the requirement that $A=\langle A\rangle=\partial
F/\partial \sigma_0$. As usual, it is assumed that in the
thermodynamic limit the relative fluctuations in the total area become
negligible, and there is no distinction between $\langle A\rangle$ and
$A$.

The renewed interest in the meaning of surface tension during the past
decade is very much linked with the rapid development in computer
modeling and simulations of bilayer membranes. In molecular
simulations the number of lipids $N$ is usually fixed, but the total
area in not. Membranes are no longer treated as incompressible thin
films, but rather as stretchable/compressible surfaces whose elastic
response result from their intermolecular forces. With this point of
view in mind, it is clear that the intrinsic tension $\sigma_0$ in
Helfrich Hamiltonian represents an elastic coefficient which,
potentially, may be related to the mechanical tension $\tau$. There
is, of course, no reason to expect that the elastic energy $E$ of the
membrane is linear in $A$ \cite{remark_added}. A quadratic elastic function
$E=1/2K_A(A-A_0)^2$ seems like a more appropriate form, where $K_A$ is
the stretching/compression modulus and $A_0$ is the relaxed area of
the membrane \cite{farago_pincus:03} (also known as Schulman's area
\cite{schulman}). Expressing the total area as the sum of the
projected and undulations areas $A=A_p+\delta A$, and assuming that
$\delta A\ll |A_p-A_0|$, yields the linear approximation $E\simeq
1/2K_A(A_p-A_0)^2+K_A(A_p-A_0)\delta A$ from which one identifies that
$\sigma_0=K_A(A_p-A_0)$. The assumption that $\delta A\ll |A_p-A_0|$
becomes increasingly accurate at high tensions; but, nevertheless, the
linear relationship between $E$ and $A$ is used in
Eq.(\ref{eq:helfhamiltonian2}) over the entire range of intrinsic
tensions.

\begin{figure}[t]
\begin{center}
\scalebox{0.55}{\centering \includegraphics{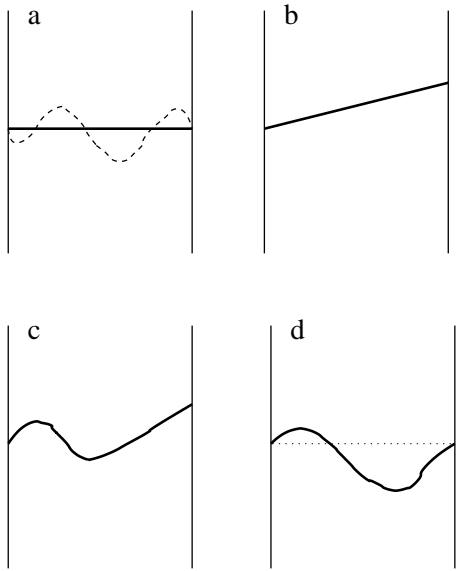}}
\end{center}
\vspace{-0.5cm}
\caption{A fluctuating membrane with a projected area $A_p$ at four
different states: (a) the reference undeformed state, (b) under shear
deformation, (c) subjected to an arbitrary deformation, and (d) as in
(c) but with periodic boundary conditions.  Each state is
characterized by the mean height profile function $\bar{h}(x,y)$ which
is indicated by a thick solid line. The dashed thin line in (a) serves
as a reminder that the membrane is fluctuating around its mean
profile.}
\label{fig:1}
\end{figure}
\section{Helfrich elastic free energy}
\label{sec:helfrichfree}

The proof that $r=\tau$ consists of three steps and involves the
introduction of two additional surface tensions: (i) $\sigma$, the
renormalized surface tension, and (ii) $\mu$, the {\em out-of-plane}\/
shear modulus (which, as it will turn out below, is also a surface
tension). Let us first understand what these two quantities
represent. The out-of-plane shear modulus $\mu$ is the force per unit
length required to introduce the deformation depicted in
fig.~\ref{fig:1}(b) from the initial reference state depicted in
fig.~\ref{fig:1}(a). This deformation is achieved by applying opposite
normal forces on boundaries of the membrane. (Notice that the
out-of-plane shear deformation changes the area of the membrane, but
preserves the volume of the three-dimensional Euclidean metric. We
will revisit this point later in section \ref{sec:rotational}.) The
shear modulus is {\em not}\/ related to the deformation of a specific
configuration but rather to the deformation of a fluctuating
membrane. The out of plane shear is a special case of a more general
family of elastic deformations resulting from the action of normal
stress fields. A representative of this family of deformations is
shown schematically in fig.~\ref{fig:1}(c). Mathematically, such
deformations can be characterized by the mean height profile of the
membrane: $\langle h(x,y) \rangle\equiv \bar{h}(x,y)$.  This function,
which vanishes in the undeformed reference state $\bar{h}(x,y) =0$,
serves as the strain field that describes arbitrary deformations of
the fluctuating membrane at fixed projected area $A_p$. Specifically
for the shear deformation shown in fig.~\ref{fig:1}(b),
$\bar{h}(x,y)=\epsilon x$ where $\epsilon$ is the shear strain. The
normal stress field associated with the strain function $\bar{h}(x,y)$
can be derived from the elastic free energy $F(\bar{h})$. Based on the
same physical arguments used to justify the introduction of Helfrich
effective Hamiltonian Eq.(\ref{eq:helfhamiltonian2}), we hereby assume
that the elastic free energy $F(\bar{h})$ can be written in a similar
form
\begin{equation}
F\left(\bar{h}\left(x,y\right)\right)=\sigma
\tilde{A}\left(\bar{h}\right)+\frac{1}{2}\kappa
(\tilde{J})^2\left(\bar{h}\right), 
\label{eq:helffree}
\end{equation}
where $\tilde{A}(\bar{h})$ and $\tilde{J}(\bar{h})$ are the total area
and integrated total curvature of the mean height profile of the
membrane, $\bar{h}(x,y)$. Our conjecture of Eq.(\ref{eq:helffree}) is
based on the fact (which will be proved in the following section
\ref{sec:lrt}) that this form of the free energy yields the
experimentally and computationally well established
Eq.(\ref{eq:q2coef}) for the spectral intensity. The coefficient
$\sigma$ and $\kappa$ appearing in Eq.(\ref{eq:helffree}) are the {\em
renormalized}\/ surface tension and bending rigidity.  These
quantities are thermodynamic properties of the membranes which, in
general, include entropic contributions and, therefore, are different
than the corresponding microscopic coefficients appearing in the
Helfrich effective Hamiltonian.

\subsection{Linear response theory}
\label{sec:lrt}

The first step in the proof that $r=\tau$ is to show that $r=\sigma$,
which follows from static linear response theory. Detailed proof of
this point can be found in ref.~\cite{farago_pincus:04}; here we
provide a short version of the derivation. Let us consider an elastic
deformation with a function $\bar{h}(x,y)$ that satisfies periodic
boundary conditions (see, e.g., fig.~\ref{fig:1}(d)). Introducing
$\bar{h}_q$, the Fourier transform of the function $\bar{h}(x,y)$, and
expanding $\tilde{A}$ and $\tilde{J}$ in powers of $\bar{h}_q$, we
find
\begin{equation}
F\left(\left\{\bar{h}_q\right\}\right)=\sigma A_p+\frac{A_p}{2}
\sum_{\vec{q}}\left[\sigma q^2+\kappa q^4\right]\bar{h}_q\bar{h}_{-q}
+{\cal O}\left(|\bar{h}_q|^4\right)
\label{eq:fourierfree}
\end{equation}
where $\bar{h}_{-q}=\bar{h}_{q}^*$ since the function $\bar{h}(x,y)$
is real \cite{remark2}.

Let us now consider the perturbed Hamiltonian ${\cal H}={\cal
H}_0-\sum_{\vec{q}}j_qh_q$, where ${\cal H}_0$ is the Hamiltonian of
the membrane. Notice that we do not need to know the specific form of
${\cal H}_0$ and, in particular, we do not assume here that ${\cal
H}_0$ is necessarily the Helfrich Hamiltonian
Eq.(\ref{eq:helfhamiltonian2}) or its quadratic approximation
Eq.(\ref{eq:mongehamiltonian}). Instead, we assume that the free
energy is given by Eq.(\ref{eq:helffree}) [or by its Fourier space
counterpart Eq.(\ref{eq:fourierfree})]; and as we shall now show, this
is the key assumption that ultimately leads to Eq.(\ref{eq:q2coef}).
%The only assumptions we make are that (i) when periodic boundary conditions are applied, ${\cal H}_0(h(x,y))$ can be expressed in terms of the Fourier transform $h_q$, i.e. as ${\cal H}_0(\{h_q\})$, and that (ii) the free energy associated with ${\cal H}$ is given by Eq.(\ref{eq:helffree}), or its Fourier space counterpart Eq.(\ref{eq:fourierfree}). The first assumption is obviously correct since ${\cal H}_0$ is some functional of $h(x,y)$ and its derivatives, and one can be simply replace them with their Fourier transformations. The second assumption will be discussed in the following sections.
The derivation is as follows: Introducing the Gibbs free energy
$G(\{j_q\})=-k_BT\ln[{\rm Tr}\,\exp(-{\cal H}/k_BT)]$, and taking the
derivative of $G$ with respect to $j_q$ yields
\begin{equation}
\bar{h}_q=-\frac{dG\left(\left\{j_q\right\}\right)}{dj_q}.
\label{eq:linres1}
\end{equation}
The conjugate thermodynamic variables $\bar{h}_q$ and $j_q$ can be
also related to each other via
\begin{equation}
j_q=\frac{dF\left(\left\{\bar{h}_q\right\}\right)}{d\bar{h}_q},
\label{eq:linres2}
\end{equation}
which, by using Eq.(\ref{eq:fourierfree}) for the Helmholtz free
energy $F$, reads
\begin{equation}
j_q=A_p\left(\sigma q^2+\kappa q^4\right) h_{-q}
+{\cal O}\left(|\bar{h}_q|^3\right).
\label{eq:linres3}
\end{equation}
Deriving Eq.(\ref{eq:linres1}) with respect to $j_{-q}$ yields the
relationship between the elastic linear response and the equilibrium
fluctuations in $h_q$
\begin{equation}
\left\langle |h_q|^2\right\rangle_0= 
k_BT\frac{d \bar{h}_{q}}{d j_{-q}}{\Biggm |}_{j_q=0}=
k_BT\left(\frac{d j_{-q}}{d \bar{h}_{q}}\right)^{-1}{\Biggm |}_{\bar{h}_q=0}
,
\label{eq:linresp4}
\end{equation}
where $\langle \cdots\rangle_0$ denote a thermal average using the
unperturbed Hamiltonian ${\cal H}_0$. Using Eq.(\ref{eq:linres3}) in
Eq.(\ref{eq:linresp4}), one arrives to the result that
 \begin{equation}
\langle |h_q|^2\rangle=\frac{k_BT}{A_p(\sigma q^2+\kappa q^4)}.
\label{eq:equipart2}
\end{equation}
This completes the proof that $r=\sigma$ which, as noted above, is
independent of the explicit form of ${\cal H}_0$.

\subsection{The renormalized tension is a shear modulus}
\label{sec:shear}

The second step is to prove that $\sigma=\mu$. This can be done easily
by considering the shear deformation $\bar{h}(x,y)=\epsilon x$ shown
in fig.~\ref{fig:1}(b). The out of plane shear modulus is defined by
the expansion of the free energy density in powers of $\epsilon$
\begin{equation}
\frac{F}{A_p} =f_0+\frac{\mu}{2}\epsilon^2+{\cal O}(\epsilon^4).
\label{eq:helfshear}
\end{equation} 
(The expansion includes only even powers of $\epsilon$ because of the
symmetry of the problem with respect to reflections in the direction
normal to the projected area.). However, for the shear deformation,
the mean profile is flat, i.e. $\tilde{J}=0$, and therefore
Eq.(\ref{eq:helffree}) reads
\begin{equation}
F\left(\bar{h}\left(x,y\right)\right)=\sigma
\tilde{A}.
\label{eq:helfshear2}
\end{equation}
The area $\tilde{A}$ and $\epsilon$ are related by
$\tilde{A}=\sqrt{1+\epsilon^2}A_p=A_p[1+\epsilon^2/2+{\cal
O}(\epsilon^4)]$, which yields
\begin{equation}
\frac{F}{A_p} =f_0+\frac{\sigma}{2}\epsilon^2+\ldots\ .
\label{eq:helfshear3}
\end{equation}
Comparison of Eqs.(\ref{eq:helfshear}) and (\ref{eq:helfshear3}) leads
to the result $\sigma=\mu$ which, again, is independent of the
explicit form of the membrane Hamiltonian ${\cal H}_0$.

\begin{figure}[t]
\begin{center}
\scalebox{0.45}{\centering \includegraphics{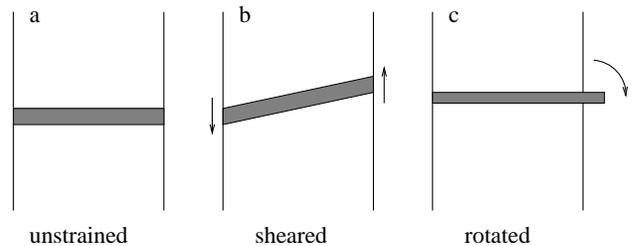}}
\end{center}
\vspace{-0.5cm}
\caption{(a) The unstrained reference state of a thin film. (b) The
film subjected to a simple shear deformation. (c) The film subjected
to a pure shear deformation. The two deformed states have the same
elastic free energy since they can be transformed into each other by
rotation. The volume of the film (indicated by the gray shaded area)
is the same at all three states.}
\label{fig:2}
\end{figure}

\subsection{Rotational invariance}
\label{sec:rotational}

Having demonstrated that $r=\sigma=\mu$, the last step in the proof
that $r=\tau$ is to show that $\tau=\sigma=\mu$. This follows
immediately from the invariance of Helfrich free energy
Eq.(\ref{eq:helffree}) with respect to rigid transformations.
Specifically, the sheared membrane which is replotted in
fig.~\ref{fig:2}(b), can be rotated (see fig.~\ref{fig:2}(c)) so that
the mean profile of the deformed membrane lies in the $(x,y)$ plane
defined by the undeformed state (fig.~\ref{fig:2}(a)). In this
orientation $\tilde{A}=A_p$, and from Eq.(\ref{eq:helfshear2}) one
gets that the free energy cost of imposing the deformation in
fig.~\ref{fig:2}(c) is $F=\sigma A_p$. From this result one finds that
the mechanical tension
\begin{equation}
\tau=\frac{dF}{dA_p}=\sigma
\label{eq:frametens1}
\end{equation}. 

While Eq.(\ref{eq:frametens1}) is correct, it misses an important
issue which cannot be captured within the framework of Helfrich model
that treats the membrane as a 2D manifold with no 3D volume. In
figs.~\ref{fig:2}(a)-(c) we have intentionally drawn the membranes as
thin films, which highlights a very important point. The transition
from the reference state (a) to the deformed state (c) is achieved via
a sequence of two volume-preserving transformations - shear followed
by rotation. Therefore, the reference state and the deformed one have
different areas but the same volume, which means that in
Eq.(\ref{eq:frametens1}) one must take the derivative at a constant
volume:
\begin{equation}
\tau=\frac{\partial F}{\partial A_p}\Biggm|_{V}=\mu.
\label{eq:frametens2}
\end{equation}
Experimentally and computationally, there is no practical way to fix
or even determine the volume of a bilayer at the molecular
resolution. This, however, does not render the above argument
irrelevant to lipid bilayer membranes. The membrane and the embedding
fluid medium are placed in a ``container'' whose volume can be easily
controlled. Eq.(\ref{eq:frametens2}) states that the frame tension of
the bilayer can be measured by changing the cross-sectional area of
the container while keeping its volume fixed. This tension is
associated with the entire ``interfacial region'' of the system that
includes both the bilayer of lipids as well as the hydration layers of
structured water around the bilayer. The fluid bulk water has no
elastic response to volume preserving deformations and, therefore, it
make no contribution to $\tau$.

The fact that $\tau$ is associated with a volume preserving
deformation is important. The reader may have noticed that in
Eq.(\ref{eq:frametens1}) we wrote $\tau=\sigma$, while in
Eq.(\ref{eq:frametens2}) we preferred the equality
$\tau=\mu$. Obviously, the latter is also correct since, as discussed
above in section \ref{sec:shear}, $\sigma=\mu$. In writing $\tau=\mu$
in Eq.(\ref{eq:frametens2}) we wish to emphasize the fact that the
frame tension is also a shear modulus.  It represents the response of
the system to ``pure shear'' deformations (fig.~\ref{fig:2}(c)), which
is the same as the response to ``simple shear''
(fig.~\ref{fig:2}(b)). The modulus of pure shear can be derived by
considering the free energy cost associated with changing the
dimensions of a system from $L_x\times L_y\times L_z$ to
$[L_x(1+\epsilon)]\times [L_y(1+\epsilon)]\times
[L_z/(1+\epsilon)^2]$. The expansion of the free energy density in
powers of $\epsilon$ reads
\begin{equation}
\frac{F}{L_xL_yL_z}=f_0+(2P_z-P_x-P_y)\epsilon+{\cal O}(\epsilon^2),
\label{eq:frametens3}
\end{equation}
where $P_i$ is the pressure along the $i$-th Cartesian axis.  Due to
the deformation, the projected area $A_p=L_xL_y$ changes by $\Delta
A_p\simeq 2A_p\epsilon$. Therefore, Eq.(\ref{eq:frametens3}) can also be
written as
\begin{equation}
F=F_0+L_z\Delta A_p\left[P_z-\frac{P_x+P_y}{2}\right]+
 {\cal O}\left(\left(\Delta A_p\right)^2\right),
\end{equation}
and the frame tension is given by 
\begin{widetext}
\begin{equation}
\tau=\lim_{\Delta A_p\rightarrow 0}\frac{F-F_0}{\Delta A_P}
=L_z\left[P_z-\frac{P_x+P_y}{2}\right]=L_z\left[P_n-P_t\right],
\label{eq:frame4}
\end{equation}
\end{widetext}
where $P_n$ and $P_t$ are, respectively, the normal and transverse
components of the pressure tensor (the negative Cauchy stress tensor)
relative to the plane of the membrane. Eq.(\ref{eq:frame4}) is known
as the mechanical definition of the surface tension
\cite{rowlinson_widom}.

\subsection{The approximated Hamiltonian} 

At this point only one question is left: why does one get $r=\sigma_0$
rather than $r=\tau$ when dealing with the quadratic approximation of
the effective surface Hamiltonian, Eq.(\ref{eq:mongehamiltonian})? The
answer is simple - the approximated Hamiltonian is not rotationally
invariant. This striking fact (which has been discussed by Grinstein
and Pelcovitz in the context of lamellar liquid crystalline phases
\cite{grinstein:81}) can be demonstrated by considering a certain
configuration parametrized by the height function $h(x,y)$ which
satisfies periodic boundary conditions, and evaluating the elastic
energy cost corresponding to simple and pure shear deformations. As
discussed above, the result that $r=\mu$ is Hamiltonian-independent,
but the following conclusion that $\mu=\tau$ depends on rotational
invariance, i.e., on the fact that the two shear deformations result
in the same elastic response. This, unfortunately, is not the case
with Eq.(\ref{eq:mongehamiltonian}). For the simple shear deformation
$h_{\mu}(x,y)=h(x,y)+\epsilon x$, and upon substituting $h_{\mu}(x,y)$
in Eq.(\ref{eq:mongehamiltonian}) one gets
\begin{widetext}
\begin{equation}
{\cal H}^M_2\left(h_{\mu}\left(x,y\right)\right)=\sigma_0 A_p +\int
dxdy\left[\frac{\sigma_0}{2} \left(\nabla h_{\mu}\right)^2
+\frac{\kappa_0}{2}\left(\nabla^2 h_{\mu}\right)^2 \right]={\cal
H}^M_2\left(h\left(x,y\right)\right)+\sigma_0A_p\frac{\epsilon^2}{2}. 
\label{eq:sshearmonge}
\end{equation}
\end{widetext}
From this result one finds that 
\begin{equation}
\mu=\frac{1}{A_p}\frac{d^2\left\langle{\cal H}^M_2
\left(h_{\mu}\left(x,y\right)\right)\right\rangle}
{d \epsilon^2}= \sigma_0, 
\label{eq:simplemonge}
\end{equation}
and, thus, $r=\mu=\sigma_0$. This conclusion that $r=\sigma_0$ is in
agreement with what the equipartition theorem Eq.(\ref{eq:equipart})
predicts for the approximated Hamiltonian. The response to pure shear
is determined by considering the transformation
$h_{\tau}(x',y')=h(x(1+\epsilon),y(1+\epsilon))/(1+\epsilon)^2$, with
$0<x'\leq L_x(1+\epsilon),\ 0<y'\leq L_y(1+\epsilon)$. For this
deformation: $A_p(\epsilon)=A_p(\epsilon=0)\cdot(1+\epsilon)^2$,
$dx'dy'=(1+\epsilon)^2dxdy$, $\nabla h_{\tau}=\nabla
h/(1+\epsilon)^3$, and $\nabla^2 h_{\tau}=\nabla^2
h/(1+\epsilon)^4$. When these relations are used in
Eq.(\ref{eq:mongehamiltonian}), we gets:
\begin{widetext}
\begin{eqnarray}
{\cal H}^M_2\left(h_{\tau}\left(x,y\right)\right)&=&(1+\epsilon)^2\sigma_0
A_p(\epsilon=0)+\int dx'dy'\left[\frac{\sigma_0}{2}
\left(\nabla h_{\tau}\right)^2 +\frac{\kappa_0}{2}\left(\nabla^2
h_{\tau}\right)^2 \right]=\\
{\cal H}^M_2\left(h\left(x,y\right)\right)
&+&2\epsilon\sigma_0 A_p(\epsilon=0)-\int dxdy\left[2\epsilon\sigma_0
\left(\nabla h\right)^2+3\epsilon\kappa_0\left(\nabla^2
h\right)^2 \right]+{\cal O}(\epsilon^2).\nonumber
\label{eq:pshearmonge}
\end{eqnarray}
From this result, one derives the mechanical tension, which is given by  
\begin{equation}
\tau=\frac{1}{2A_p}\frac{d\left\langle{\cal H}^M_2
\left(h_{\tau}\left(x,y\right)\right)\right\rangle} 
{d \epsilon}\Biggm|_{\epsilon=0}=
\sigma_0-\frac{1}{A_p}\left\langle\int dxdy\left[\sigma_0
\left(\nabla h\right)^2+\frac{3}{2}\kappa_0\left(\nabla^2
h\right)^2 \right]\right\rangle.
\label{eq:puremonge}
\end{equation}
\end{widetext}
The second term on the right hand side of Eq.(\ref{eq:puremonge}) is
the entropic part of the mechanical tension, evaluated within the
framework of the approximated quadratic Hamiltonian. This entropic
contribution to $\tau$ is not reflected in the fluctuation spectrum of
the quadratic Hamiltonian, which has the non-physical feature of not
being rotationally invariant.

\section{Computer simulations}

\begin{figure}[t]
\begin{center}
\scalebox{0.5}{\centering \includegraphics{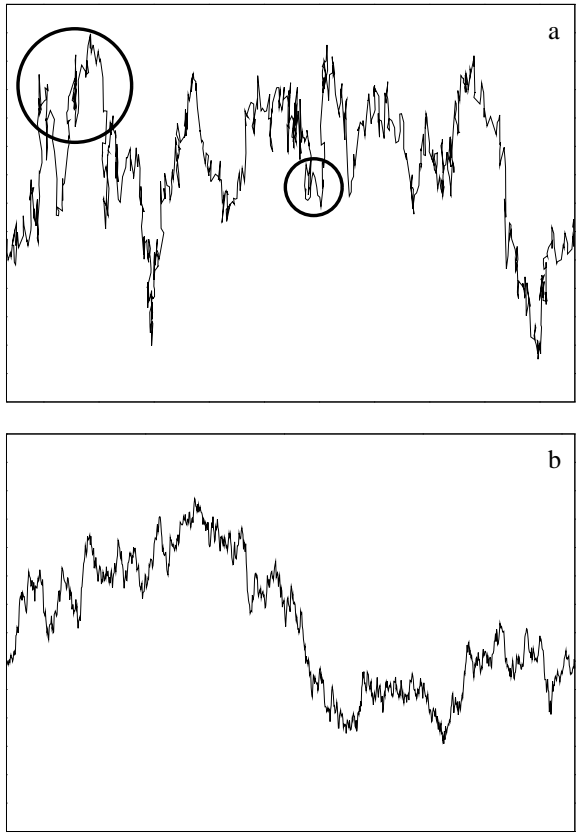}}
\end{center}
\vspace{-0.5cm}
\caption{Typical configurations of the chain in (a) the rotationally
invariant Hamiltonian Eq.(\ref{eq:helf1d}), and (b) its quadratic
approximation Eq.(\ref{eq:monge1d}). Both configurations have been
obtained for $\sigma_0=0.75 k_BT/l_p$ and $\kappa_0=0$. In (a) the
points of the chain are allowed be anywhere in the available 2D space
which creates overhangs such as those shown inside the bold
circles. In (b) the points move only vertically to the line that
connects the end points of the chain.}
\label{fig:3}
\end{figure}

To summarize our discussion: For any Hamiltonian whose corresponding
free energy is given by the rotationally invariant free energy
Eq.(\ref{eq:helffree}), the correct result is $r=\tau$. For the not
rotationally invariant quadratic Hamiltonian
Eq.(\ref{eq:mongehamiltonian}), $r=\sigma_0\neq\tau$. In order to test
these predictions, we performed Monte Carlo simulations of the
one-dimensional (1D) analogs of Eqs.(\ref{eq:helfhamiltonian2}) and
(\ref{eq:mongehamiltonian}). Within these two models, the membrane is
represented by a string of $N=1024$ points, the positions of which in
2D space are given by $\{\vec{r}_i=(x_i,h_i)\}$. In both cases, the
simulations are performed at a constant projected length $L_p$ with
periodic boundary conditions. Denoting by
$\vec{b}_i=\vec{r}_{i+1}-\vec{r}_i$, the distance vector (``bond'')
between adjacent points, the 1D analog of the rotationally invariant
Helfrich effective Hamiltonian is given by \cite{remark3}
\begin{equation}
{\cal H}_1=\sigma_0\sum_i|\vec{b}_i|+\kappa_0\sum_i\left[1
-\frac{\vec{b}_i\cdot\vec{b}_{i-1}}{|\vec{b}_i||\vec{b}_{i-1}|}\right].
\label{eq:helf1d}
\end{equation}
The quadratic approximation of this Hamiltonian is given by
\begin{widetext}
\begin{equation}
{\cal H}_2=\sigma_0L_p+\frac{\sigma_0}{2l_p}
\sum_i\left(h_{i+1}-h_{i}\right)^2+
\frac{\kappa_0}{2l_p^2}
\sum_i\left(h_{i+1}+h_{i-1}-2h_i\right)^2,
\label{eq:monge1d}
\end{equation}
\end{widetext}
where $l_p=L_P/N$. Notice that in the 1D models, the tension has units
of a force (energy per unit length). A major difference between the
two 1D models is related to the positions of the points. In the
rotationally invariant case Eq.(\ref{eq:helf1d}), the points are
allowed to be anywhere in the available 2D space. At low tensions,
this creates configurations in which the chain forms overhangs (see
e.g., fig.~\ref{fig:3}(a)). The existence of such configurations does
not invalidate any of the above discussion which only requires that
the mean height profile $\bar{h}(x,y)$ is a well defined
function. Within the quadratic approximation Eq.(\ref{eq:monge1d}),
the points are allowed to move only in the direction normal to the
projected length [in the spirit of Eq.(\ref{eq:mongehamiltonian}) in
which the height is measured from the $x-y$ plane]. Thus, the position
of the $i$-th point is given by $\vec{r}_i=(il_p,h_i)$ and, obviously,
such moves do not generate any overhangs (see a typical configuration
in fig.~\ref{fig:3}(b)).

In the simulations, we vary $\sigma_0$ and measure both the mechanical
tension, $\tau$, and the $q^2$-coefficient, $r$. For each value of
$\sigma_0$, the simulations extended over $2-4\times 10^8$ MC time
units, where each time unit consists of $N$ single particle move
attempts and one collective ``mode excitation Monte Carlo'' (MEMC)
move that accelerates the very slow relaxation dynamics of the 10
largest Fourier modes \cite{memc}. The introduction of MEMC moves is
essential for the equilibration of the system. The mechanical tension
is calculated using the 1D equivalent of Eq.(\ref{eq:frame4}),
$\tau=f_n-f_t$, where $f_n$ and $f_t$ are the normal and transverse
forces acting of the chain. Expressing these forces as thermal
averages, one arrives to the following virial formulae. For the
rotationally invariant Helfrich Hamiltonian ${\cal H}_1$
\begin{widetext}
\begin{eqnarray}
\tau&=&\left\langle \sum_i\left[\sigma_0\frac{\left(x_{i+1}-x_i\right)^2-
\left(h_{i+1}-h_i\right)^2}{l_p|\vec{b}_i|}\right.\right.\nonumber \\
&+&\kappa_0\frac{2}{l_p|\vec{b}_{i-1}||\vec{b}_i|}
\left(
\frac{\left(h_{i}-h_{i-1}\right)\left(h_{i+1}-h_i\right)
\left(x_i-x_{i-1}\right)^2-\left(x_{i}-x_{i-1}\right)\left(x_{i+1}-x_i\right)
\left(h_i-h_{i-1}\right)^2}{|\vec{b}_{i-1}|^2}\right.\nonumber\\
&+&\left.\left.\left.\frac
{\left(x_{i}-x_{i-1}\right)^2+\left(h_{i}-h_{i-1}\right)^2}{|\vec{b}_i|^2}
\right)\right]\right\rangle.
\label{eq:taufull}
\end{eqnarray}
For the approximated quadratic Hamiltonian ${\cal H}_2$
\begin{eqnarray}
\tau&=&\sigma_0-\left\langle\sum_i\left[\frac{3\sigma_0}{2l_p^2}
\left(h_{i+1}-h_{i}\right)^2+\frac{2\kappa_0}{l_p^3}\left(h_{i+1}+h_{i-1}-2h_i
\right)^2
\right]\right\rangle.
\label{eq:taumonge}
\end{eqnarray}
\end{widetext}
The $q^2$-coefficient $r$ is obtained by taking the Fourier transform
of the the function $h_i$,
%(with the discrete variable $i$, and not the
%physical coordinate $x_i$, playing the role of the coordinate along
%the end-to-end direction) 
and fitting the measurements to Eq.(\ref{eq:q2coef}). Our results for
$r$ are based on the analysis of the fluctuations of the 10 largest
Fourier modes.

\begin{figure}[t]
\begin{center}
\scalebox{0.375}{\centering \includegraphics{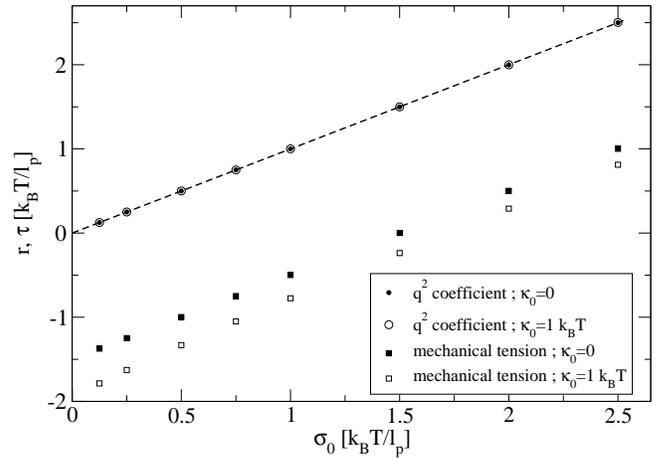}}
\end{center}
\vspace{-0.5cm}
\caption{Simulation results for Hamiltonian ${\cal H}_2$
[Eq.(\ref{eq:monge1d})]: The $q^2$-coefficient $r$ (circles) and the
mechanical tension $\tau$ (squares) as a function of the intrinsic
tension $\sigma_0$. Results are shown for both $\kappa_0=0$ (solid
symbols) and $\kappa_0=1k_BT$ (open symbols). The dashed line is a
guide to the eye for the relationship $r=\sigma_0$.}
\label{fig:4}
\end{figure}

\begin{figure}[t]
\begin{center}
\scalebox{0.375}{\centering \includegraphics{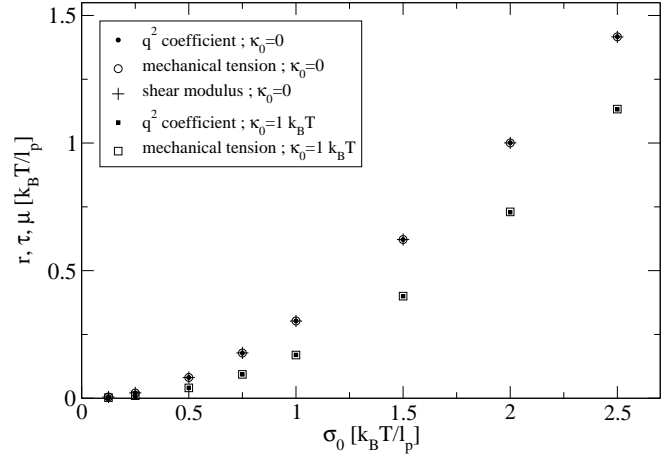}}
\end{center}
\vspace{-0.5cm}
\caption{Simulation results for Hamiltonian ${\cal H}_1$
[Eq.(\ref{eq:helf1d})]: The $q^2$-coefficient $r$ (solid symbols) and
the mechanical tension $\tau$ (open symbols) as a function of the
intrinsic tension $\sigma_0$. Results are shown for both $\kappa_0=0$
(circles) and $\kappa_0=1k_BT$ (squares). The pluses denote our
results for the shear modulus $\mu$ for $\kappa_0=0$.}
\label{fig:5}
\end{figure}

Our simulation results for the quadratic Hamiltonian ${\cal H}_2$ are
summarized in fig.~\ref{fig:4}. For both $\kappa_0=0$ and
$\kappa_0=1k_BT$ and for all values of $\sigma_0$, we find that,
indeed, all our measurements of the $q^2$-coefficient agree with the
predicted relationship $r=\sigma_0$ (which is denoted by the dashed
line). As expected from Eq.(\ref{eq:taumonge}), the mechanical tension
$\tau$ is smaller than $r$ and for small values of $\sigma_0$ even
gets negative values. In comparison, the simulation results for the
rotationally invariant Helfrich Hamiltonian ${\cal H}_1$ are shown in
fig~\ref{fig:5}. In agreement with our expectations for this case, the
various tensions satisfy the relationship that
$r=\tau\neq\sigma_0$. To further demonstrate the validity of our
discussion in section \ref{sec:helfrichfree}, we computed the simple
shear modulus $\mu$, which for Hamiltonian ${\cal H}_1$, is also
expected to be equal to $r$ and $\tau$.  The shear modulus $\mu$ can
be computed using a rather cumbersome virial expression. For
$\kappa_0=0$, the virial expression for $\mu$ takes the more simple
form:
\begin{widetext}
\begin{equation}
\mu=\left\langle\sum_i\sigma_0\frac
{\left(x_{i+1}-x_i\right)^4}{l_p|\vec{b}_i|^3}
-\frac{1}{l_pk_BT}\left(\sum_i\sigma_0\frac{
\left(x_{i+1}-x_i\right)\left(h_{i+1}-h_i\right)}{|\vec{b}_i|}\right)^2
\right\rangle.
\label{eq:mufull}
\end{equation}
\end{widetext}
Our results for $\mu$ agree perfectly with the results for $\tau$ and
$r$.


\section{Conclusions}

Motivated by the still ongoing debate about the various meanings of
the term ``surface tension'' in membranes, we presented here a
detailed discussion which highlights the importance of distinguishing
between Helfrich free energy and Helfrich Hamiltonian, and between the
Hamiltonian and its approximated quadratic form. Our key findings are
the followings:
\begin{enumerate}
\item We have demonstrated that, contrary to common perception, the
observation that the spectrum of thermal fluctuations follows
Eq.(\ref{eq:q2coef}), is not an evidence for the validity and accuracy
of the quadratic Helfrich Hamiltonian
Eq.(\ref{eq:mongehamiltonian}). Instead, we have derived
Eq.(\ref{eq:q2coef}) based on the assumption that the thermodynamic
properties of the membrane are correctly depicted by Heflrich free
energy Eq.(\ref{eq:helffree}).
\item In most thermodynamic theories the state of the membrane is
specified by the extensive variables $A_p$ and $A$ (or $N$)
\cite{remark4}. In our approach, Eq.(\ref{eq:helffree}) is understood
as the elastic free energy {\em functional}\/ that depends on the mean
profile around which the membrane fluctuates. Thus, the state of the
membrane should be described by not only two variables, but through a
function $\bar{h}$ that serves as the strain field for the deformed
membrane.
\item The $q^2$-coefficient is equal to the renormalized surface
tension $\sigma$, i.e., the coefficient of proportionality between the
elastic free energy and the total area of the mean profile in
Eq.(\ref{eq:helffree}). The renormalized tension is a thermodynamic
quantity that also depends on the entropy of the membrane.
\item The result that $r=\sigma$ follows from linear response theory
and, therefore, is independent of the particular form the membrane
Hamiltonian. In contrast, the conclusion that $r=\tau$ is based on the
assumption that the system is rotationally invariant and respond
equally to all shear deformations irrespective of the relative
orientations of the deformed and undeformed membranes.
\item The last point (rotational invariance) is not satisfied when the
membrane is described by the approximated quadratic form
Eq.(\ref{eq:mongehamiltonian}), which calls for a reconsideration of
some of the results derived through this model. Specifically, in
refs.~\cite{imparato:06,fournier:08,farago_pincus:03} the quadratic
form has been used for a derivation of negative mechanical tension for
positive intrinsic tension [see also Eq.(\ref{eq:puremonge}) in the
present paper]. Our computational results demonstrate that this result
is achieved only with the faulty (non-physical) quadratic Hamiltonian
${\cal H}_2$ (see fig.~\ref{fig:4}); but for the corresponding
rotationally invariant Hamiltonian ${\cal H}_1$, the mechanical
tension is always positive for $\sigma_0>0$ (see
fig.~\ref{fig:5}). Our computational results actually suggest that
$\sigma_0$ and $\tau$ vanish simultaneously.
\item To express the last point in a different manner - in order to
correctly describe the elastic properties of membranes, one needs to
include anharmonic terms in the Hamiltonian, which leads to mode
coupling.
\end{enumerate}

Acknowledgments: I am deeply grateful to Haim Diamant for numerous
stimulating discussions and for reading the manuscript with great
scrutiny. I also thank the comments of Phil Pincus and Adrian
V. Parsegian.

%----------------------------------------------------------- 
%References
%----------------------------------------------------------- 


\begin{thebibliography}{99}
\bibitem{remark1} Later, it will become clear that one needs to take
this derivative while keeping the volume of the system
fixed. Mathematically, this implies that the surface in Helfrich
Hamiltonian should be treated as a thin film whose width becomes
vanishingly small.
\bibitem{remark_added} The total elastic energy $E$ does not depend on
the total area $A$ only, but also on the distribution of the lipids
within the given area. The local density of the lipids defines the
local strain field, and $E$ is calculated by integrating the local
elastic energy density ${\cal E}$: $E=\int_A {\cal E}dS$. When
speaking about the area-dependent energy, $E(A)$, we refer to the mean
energy averaged over density fluctuations.
\bibitem{remark3} The second term in Hamiltonian (\ref{eq:helf1d})
differs from the curvature term in Eq.(13) in ref.~{\protect
\cite{fournier:08}}. This change has been introduced to avoid the
buckling instability occurring at low tensions when $L$ is allowed
fluctuate.
\end{thebibliography}
\end{document}